\documentclass[showpacs,amsmath,reprint,showkeys,pra]{revtex4-1}
\usepackage{dcolumn}    % Align table columns on decimal point
\usepackage{bm}         % bold math
\usepackage{ifpdf}
\usepackage{amssymb,lineno,amsfonts}
\usepackage{graphicx}   % Include figure files
\usepackage{bbm}
\usepackage{mathrsfs}
\usepackage{upgreek}
\usepackage{epstopdf}
\usepackage{setspace}
\usepackage{hyperref}
\usepackage[matrix,frame,arrow]{xypic}
\usepackage{float}
\usepackage{soul}
 \usepackage{tikz}
\usepackage{natbib}
\usepackage{color} 
\usepackage{subfigure}
\definecolor{med-blue}{RGB}{25,25,112} 
\hypersetup{colorlinks, linkcolor={blue},citecolor={blue}, urlcolor={black}}
\newcommand{\abs}[1]{\vert{#1}\vert}

\newcommand{\tr}{\mathrm{Tr}}

\begin{document}
	\title{Experimental emulation of  quantum non-Markovian dynamics and \\ coherence protection in the presence of information backflow }
	\author{Deepak Khurana}
	\email{deepak.khurana@students.iiserpune.ac.in}
	\author {Bijay Kumar Agarwalla}
			\email{bijay@iiserpune.ac.in }
	\author{T. S. Mahesh}
		\email{mahesh.ts@iiserpune.ac.in}
	\affiliation{ Department of Physics,
		Indian Institute of Science Education and Research, Pune 411008, India}

	\begin{abstract}
		{We experimentally emulate, in a controlled fashion, the non-Markovian dynamics of a pure dephasing spin-boson model at zero temperature. Specifically, we use a  randomized set of external radio-frequency fields to engineer a desired noise power-spectrum to effectively realize a non-Markovian environment for a single NMR qubit. The information backflow, characteristic to the non-Markovianity,  is captured in the nonmonotonicity of the decoherence function and von Neumann entropy of the system. Using such emulated non-Markovian environments, we experimentally study the efficiency of the Carr-Purcell-Meiboom-Gill dynamical decoupling (DD) sequence to inhibit the loss of coherence. Using the filter function formalism, we design optimized DD sequences that maximize coherence protection for non-Markovian environments and study their efficiencies experimentally. Finally, we discuss DD-assisted tuning of the effective non-Markovianity.    

			}
	\end{abstract}
		

\maketitle
\noindent
\section{Introduction}
 Despite promising to outperform their classical counterparts by miles, quantum technologies are inherently plagued by the inevitable interactions with the surrounding environment leading to decoherence \cite{zurek2003decoherence}, which limits their utilization to the full potential. At microscopic levels,  under certain assumptions, namely weak system-environment coupling, uncorrelated initial states, and short environmental correlation times, reduced dynamics of the open quantum system (OQS) can be described by a master equation of the Lindblad structure with a constant Lindblad operator and positive decay rates  \cite{gorini1976completely,lindblad1976generators}. Such dynamics is generally labeled as Markovian and the corresponding dynamical map describing the evolution of the reduced OQS satisfies semi-group property.
 
  In realistic scenarios,  memory effects associated with the environment must be taken into consideration when these assumptions are not justified and as a consequence the corresponding  dynamical map does not satisfy semi-group property.  In this regard, several definitions of non-Markovianity  exist in literature \cite{breuer2016colloquium,de2017dynamics,li2018concepts,rivas2014quantum,li2018concepts}. However, there is no general agreement on one definition and therefore  characterization of non-Markovian dynamics is highly context dependent \cite{li2018concepts}. Moreover, these memory effects have been shown to be resources for certain quantum information tasks \cite{huelga2012non,vasile2011continuous,bylicka2014non,chin2012quantum,dong2018non,laine2014nonlocal}, which further motivates us  to understand the deviation of OQS dynamics from a Markovian description. To this end, recently several definitions and measures to quantify the degree of non-Markovianity have emerged \cite{breuer2009measure,rivas2010entanglement,chruscinski2014degree,torre2015non,luo2012quantifying,lu2010quantum,lorenzo2013geometrical,vasile2011quantifying} from a quantum information perspective. These definitions mainly rely on studying the time evolution of certain information-theoretical   quantities  under the action of the dynamical map. In this work, we use the Breuer-Laine-Pilo (BLP) measure \cite{breuer2009measure} which is based on distinguishability (trace distance \cite{nielsen2010quantum}) between quantum states of the OQS. Distinguishability is always contractive under the action of the Markovian dynamical map. Therefore, a momentary increase in distinguishability is a signature of non-Markovianity and physically interpreted as information backflow from the environment. A comparative study of these measures for the OQS model considered in this paper, namely  pure dephasing \cite{reina2002decoherence} of a qubit, is carried out in Refs. \cite{addis2014comparative,guarnieri2014quantum}.
  All common definitions of non-Markovianity coincide in this case \cite{zeng2011equivalence,breuer2016colloquium}. However corresponding  measures  proposed to quantify the amount of non-Markovianity are not equivalent \cite{addis2014comparative,guarnieri2014quantum}.
 
Any rescue strategy developed to counter the detrimental impact of decoherence on quantum information must be quantitatively benchmarked to ensure its robustness against various kinds of realistic environments. In this regard, the non-Markovianity measure defined from a quantum information perspective enables one to study the impact of information backflow on quantum control protocols in a more quantitative manner \cite{mukherjee2015efficiency,addis2016problem,mangaud2018non,tai2014optimal,addis2015dynamical,nmirkin,reich2015exploiting}. Specifically, dynamical decoupling (DD) \cite{viola1998dynamical,viola1999dynamical} is one of the most successful techniques developed in the past two decades. Efficiency of a DD sequence is related to correlation times of the environment and accordingly various DD schemes have been designed to suit the type of the environment  \cite{khodjasteh2005fault,pasini2010optimized,uhrig2007keeping,uhrig2010rigorous,yang2008universality,yang2011preserving,viola1998dynamical,uhrig2007keeping,cywinski2008enhance, biercuk2009optimized,du2009preserving,alvarez2010performance,ajoy2011optimal,souza2011robust}. Therefore it is imperative to experimentally investigate the impact of memory effects quantified using non-Markovianity measures on the efficiency of DD sequences.

  To ensure a faithful benchmarking one has to reproduce the effect of the environment (mainly noise spectral density) in a controlled fashion. In this regard, a systematic transition from Markovian to non-Markovian dephasing dynamics was demonstrated in photonic systems \cite{liu2011experimental}. However, a full control over synthesis of noise spectral density is required for quantum control benchmarking purposes which can be achieved using artificially engineered environments \cite{soare2014experimental,ball2016effect,soare2014experimental1,edmunds2017measuring,mavadia2018experimental,zhen2016optimal}. 

Equipped with an elaborate control on quantum dynamics, nuclear magnetic resonance (NMR) systems are excellent test bed for these kind of studies. 
   In this work, using $^{1}$H nuclear spins of water molecules in a liquid-state NMR setup as a qubit-ensemble, we experimentally mimic the non-Markovian dynamics of a pure dephasing quantum spin-boson model via injection of classically colored noise. We utilize amplitude and phase-modulated external radio-frequency (RF) fields to produce a desired noise power spectrum \cite{soare2014experimental}. The signature of non-Markovianity is captured in terms of non-monotonicity of the decay of trace distance  \cite{breuer2009measure} (BLP measure) and the behavior of von Neumann entropy of reduced OQS. Using engineered non-Markovian environments, we  experimentally investigate the efficiency of Carr-Purcell-Meiboom-Gill (CPMG) DD sequence   \cite{meiboom1958modified,carr1954effects} in protecting coherence. Further, using filter function formalism \cite{biercuk2009optimized}, we design  optimized DD sequences that achieve a superior coherence protection  for a given non-Markovian environment and study their experimental efficiency. We also indicate the potential application of DD sequences in tuning the  effective non-Markovianity. 

\section{Emulation of  non-Markovian dephasing dynamics} 
\label{1}
In this work, we consider the spin-boson pure dephasing Hamiltonian \cite{reina2002decoherence}
\begin{equation}
\mathcal{H} = \omega_0 \sigma_z/2 + \sum_{k}\omega_k b_k^\dagger b_k+ \sum_{k} \sigma_z (g_kb_k + g_k^* b_k^\dagger),
\label{eq1}
\end{equation}
 consisting of precession of a single-qubit with a frequency $\omega_0$  and Pauli $z$-operator $\sigma_z$ (first term), a bosonic environment with creation (annihilation) operator $b_k^{\dagger}(b_k)$ (second term) and the mutual interaction with coupling constant $g_k$ (third term).
This Hamiltonian model is exactly solvable and leads to the decay of coherences without affecting the populations. The decoherence function is of the form $\Gamma_0(t) = e^{-\chi_0(t)}$,
\begin{equation}
\label{chifull}
\chi_0(t) = 2\int_{0}^{\infty} d\omega J(\omega) \coth (\omega/2 k_B T) \frac{\sin^2(\omega t/2)}{\omega^2},
\end{equation}
where $k_B T$ and $J(\omega) = \sum_{k} |g_k|^2 \delta(\omega-\omega_k)$ describe the thermal energy and the spectral density of the environment respectively. In the absence of initial system-environment correlations, the reduced density matrix of OQS in the interaction picture follows a master equation with a single Lindblad operator
$$\dot{\rho} = \Phi_t \rho =  \gamma_0(t)[\sigma_z\rho(t)\sigma_z-\rho(t)],\ \ \gamma_0(t) = -\dot{\Gamma}_0(t)/{\Gamma}_0(t).$$
The dynamical map $\Phi_t$ leads to non-Markovian dynamics as the decay rate $\gamma_0(t)$ becomes negative for some $t\geq 0$ depending on the temperature and spectral density of the environment. This holds according to all common definitions of non-Markovianity \cite{zeng2011equivalence,breuer2016colloquium} as pointed out earlier. However various measures of non-Markovianity are not equivalent \cite{addis2014comparative,guarnieri2014quantum}. Here we use the one based on the contractive property of trace distance $$\mathcal{D}(\rho_1,\rho_2) = ||\rho_1-\rho_2||/2,$$
where $||\rho|| = \tr(\sqrt{\rho^\dagger\rho})$ \cite{nielsen2010quantum}. Under a Markovian dynamical map $\Phi:\rho(0)\rightarrow \rho(t)$, the trace distance is always contractive, i.e.,  $$\mathcal{D}(\rho_1 (t),\rho_2 (t))\leq \mathcal{D}(\rho_1 (0),\rho_2 (0))$$
 for all pairs of initial states $\{\rho_1(0),\rho_2(0)\}$. Here, the equality holds for the evolution under a unitary map. On the other hand, a dynamical map is non-Markovian if there exists a pair of initial states for which  the trace distance shows a non-monotonic behavior. Such a non-monotonicity of the trace distance is associated with information backflow from the environment \cite{breuer2016colloquium,breuer2009measure,de2017dynamics}.     Accordingly, the BLP measure \cite{breuer2009measure} of non-Markovianity is defined as
\begin{equation*}
\mathcal{N} = \max_{\rho_1(0),\rho_2(0)} \int\displaylimits_{\sigma>0} \sigma(t) dt,~
\mbox{where,}~ \sigma (t) = \mathcal{\dot{D}}(\rho_1 (t),\rho_2 (t)).
\end{equation*}
%where $$.\\ 

%  has all the information regarding the environment relevant for reduced system dynamics. 
  In this case non-Markovianity measure $\mathcal{N}$ is maximized for any pair of antipodal initial states on the Bloch sphere \cite{wissmann2012optimal} and $\sigma (t) = \dot{\Gamma}_0 (t)$. Accordingly, non-Markovianity measure takes a simple form, $$\mathcal{N} = \sum_{k} [\Gamma_0 (t_k^f) - \Gamma_0(t_k^i)],$$ considering all the intervals $[t_k^i,t_k^f]$ wherein $\dot{\Gamma}_0 (t) > 0$.

  The reduced dynamics of the OQS under the spin-boson dephasing Hamiltonian (Eq. \ref{eq1}) can be emulated by considering its semi-classical limit  $[\omega_0 + \xi (t)]\sigma_z/2,$ where $\xi (t)$ is a stationary Gaussian stochastic process with zero mean and with a correlation function $\langle \xi(t_1) \xi(t_2)\rangle = g(t_1\!-\!t_2)$.
  The Fourier transform $S(\omega)$ of the time averaged $g(t)$ is called the noise power spectrum, which  replaces $\pi J(\omega) \coth (\omega/2k_B T)$ in Eq.~\ref{chifull}, so that the decoherence function in this limit reduces to 
 \begin{equation}
 \label{chicl}
 \chi_0^c(t) = \frac{1}{2\pi}\int_{0}^{\infty} d\omega S(\omega)\vert F_0(\omega,t) \vert^2,
 \end{equation}
 where the free-evolution filter-function $\abs{F_0(\omega,t)}^2 = 4\sin^2(\omega t/2)/\omega^2$.
 %Since $\coth (\omega/2k_B T)\rightarrow 1$ as $T\rightarrow 0$, $\chi_0(t)$  and $\chi_0^c(t)$ have similar functional forms at low temperature limit.
\begin{figure}
	\includegraphics[trim = 5.3cm 6cm 3.8cm 4.5cm, clip, width=8.5cm ]{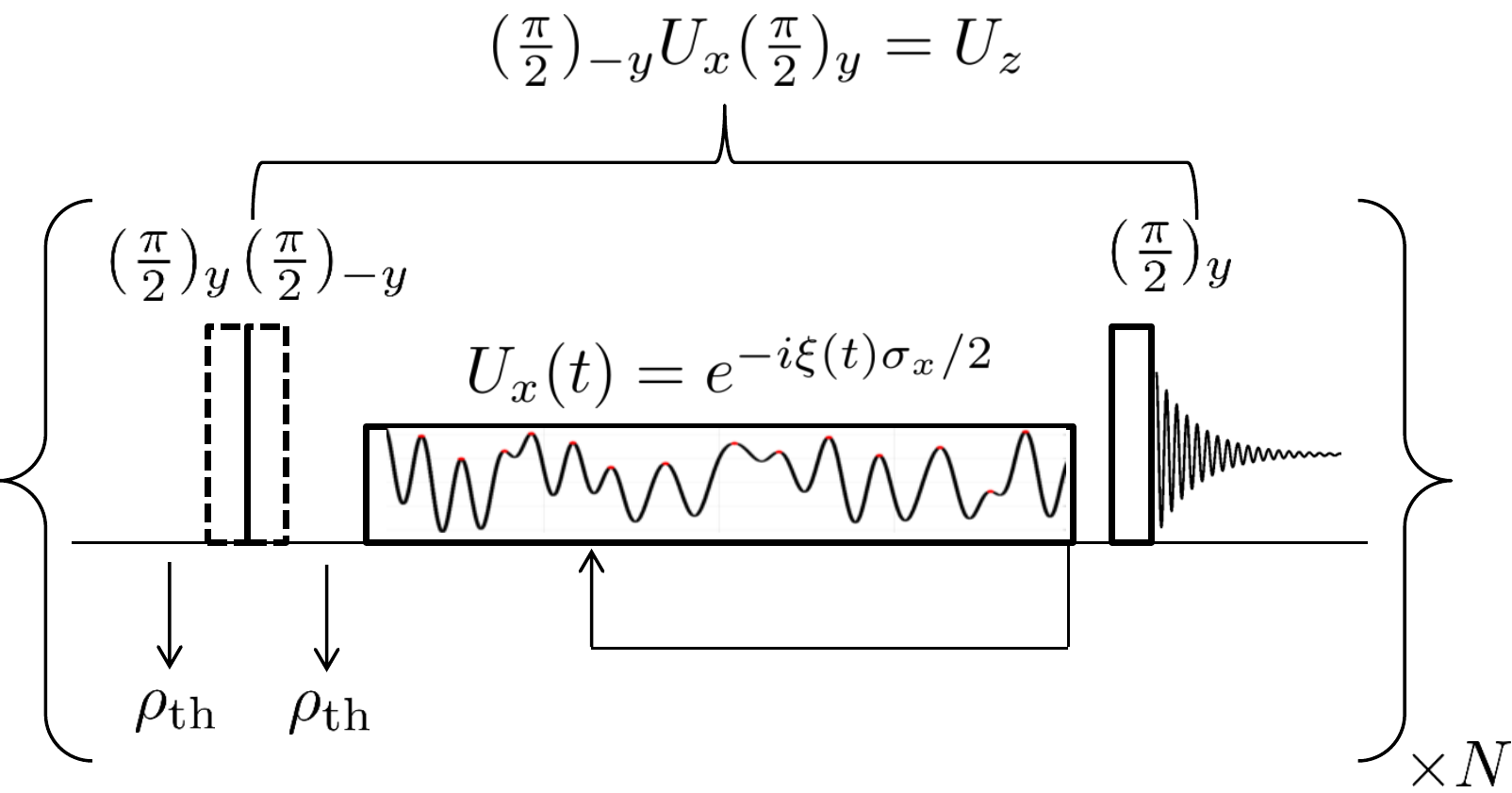}
	\caption{NMR pulse-sequence used to implement a non-Markovian dephasing dynamics. Here $\rho_{\mathrm{th}} = (\mathbb{I} + \epsilon \sigma_z)/2$ is the initial thermal state with purity factor $\epsilon$.  The final signal is obtained by averaging over $N$ independent realizations of the stochastic process $\xi(t)$ as in Eq. \ref{eng}.}
	\label{circuit}
\end{figure} 

This suggests that we can mimic the non-Markovian dynamics of a single qubit coupled to a bosonic environment with a synthetic noise power spectrum. We engineer such a power spectrum via a temporal average over a set of stochastic fields of the form,
\begin{equation}
\label{eng}
\xi (t) = \gamma \sum_{k = 1}^{M} a(k) \cos (k\omega_b t + \phi ),
\end{equation}
where $\gamma$ is strength of noise, $a(k)$  is the amplitude of the $k$th Fourier component, $\omega_b$ is the base frequency,  and $\phi \in [-\pi,\pi]$ is a random number with a uniform distribution. The resulting noise power spectrum is of the form \cite{soare2014experimental}
\begin{equation}
S(\omega) = \frac{\pi \gamma^2}{2} \sum_{k = 1}^{M} a^2(k) [\delta(\omega - k\omega_b) + \delta(\omega + k\omega_b)].
\label{somega}
\end{equation}
In our experiments, we consider the spectral density 
\begin{equation}
J(\omega) = \lambda \exp(-\omega/\omega_c)~\omega^s/\omega_c^{s-1} ,
\label{johmic}
\end{equation}
where the Ohmicity parameter $s = 1$, $s<1$ and $s>1$ corresponds to the Ohmic, sub-Ohmic and super-Ohmic spectrum, respectively. 
Comparing Eqs. \ref{chifull}, \ref{johmic}, and \ref{somega}, we find that for the spectral density mentioned above,
\begin{equation}
a^2(k) = \frac{(k \omega_b)^s }{\omega_c^{s-1}}e^{-k\omega_b/\omega_c}\coth\left(\frac{k\omega_b}{2k_B T}\right), \ \ \ \gamma^2 = 2\lambda,
\end{equation} 
which implies that we can emulate pure dephasing dynamics with a non-Markovian behavior of a bosonic reservoir by properly tuning $s$, $\gamma$, $T$, and $\omega_c$. In this work, we confine ourselves to the zero-temperature ($T= 0$) case.

The functional forms of $\mathcal{N}$ 
versus the dimensionless coupling constant $\lambda$ and Ohmicity parameter $s$ have been investigated in \cite{addis2014comparative,guarnieri2014quantum}. It has been shown in earlier works \cite{addis2014comparative,guarnieri2014quantum,addis2015dynamical} that dephasing dynamics becomes BLP non-Markovian ($\mathcal{N}> 0$) for $s>2$ and $T = 0$. 
Interestingly, environments corresponding to $s\leq2$  do not give rise to non-Markovianity, according to the BLP measure, despite having non-zero correlation times.
\begin{figure}[h]
	\includegraphics[trim = 4.2cm 7.2cm 1.8cm 6cm, clip, width=8.5cm ]{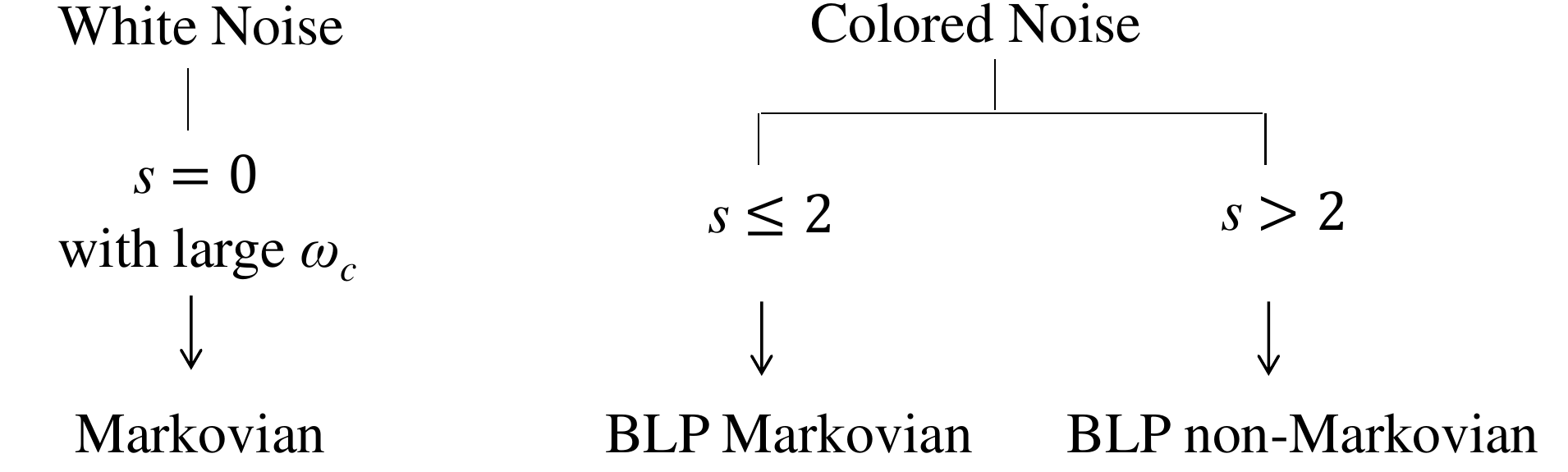}
\end{figure} 

 To experimentally emulate the non-Markovian dynamics, we take a NMR sample consisting of 20\% H$_2$O in 80\% D$_2$O, with a trace of CuSO$_4$ that shortens $^1$H longitudinal and transverse relaxation times to $T_1\approx$ 200ms and $T_2\approx$ 180ms respectively. The experiments are carried out in a Bruker 500 MHz NMR spectrometer at an ambient temperature of 300 K.
Here the two Zeeman levels of the spin-1/2 $^1$H nucleus form the qubit and stochastic controls required to engineer a desired $S(\omega)$ are realized by transverse radio-frequency (RF) fields whose amplitude and phase are modulated according to Eq. \ref{eng}.  
The corresponding NMR pulse-sequence requires an initial $(\pi/2)_y$ pulse to prepare coherence followed by a stochastic longitudinal control field $U_z(t)=e^{-i\xi(t)\sigma_z/2}$.
However, as illustrated in Fig. \ref{circuit}, $U_z(t)$ is implemented by a transverse stochastic unitary $U_x(t)$ sandwiched between $(\pi/2)_{-y}$ and $(\pi/2)_y$ pulses, wherein the $(\pi/2)_{-y}$ pulse is nullified with the initial $(\pi/2)_y$ pulse.  Finally, an effective dephasing dynamics is achieved by temporally averaging NMR signals over $N=1000$ independent realizations of the stochastic process $\xi(t)$ (Eq. \ref{eng}) consisting of $M = 1000$ Fourier components.
 \begin{figure}
	\includegraphics[trim = 3cm 1cm 3.3cm 1.1cm, clip, width=8.5cm ]{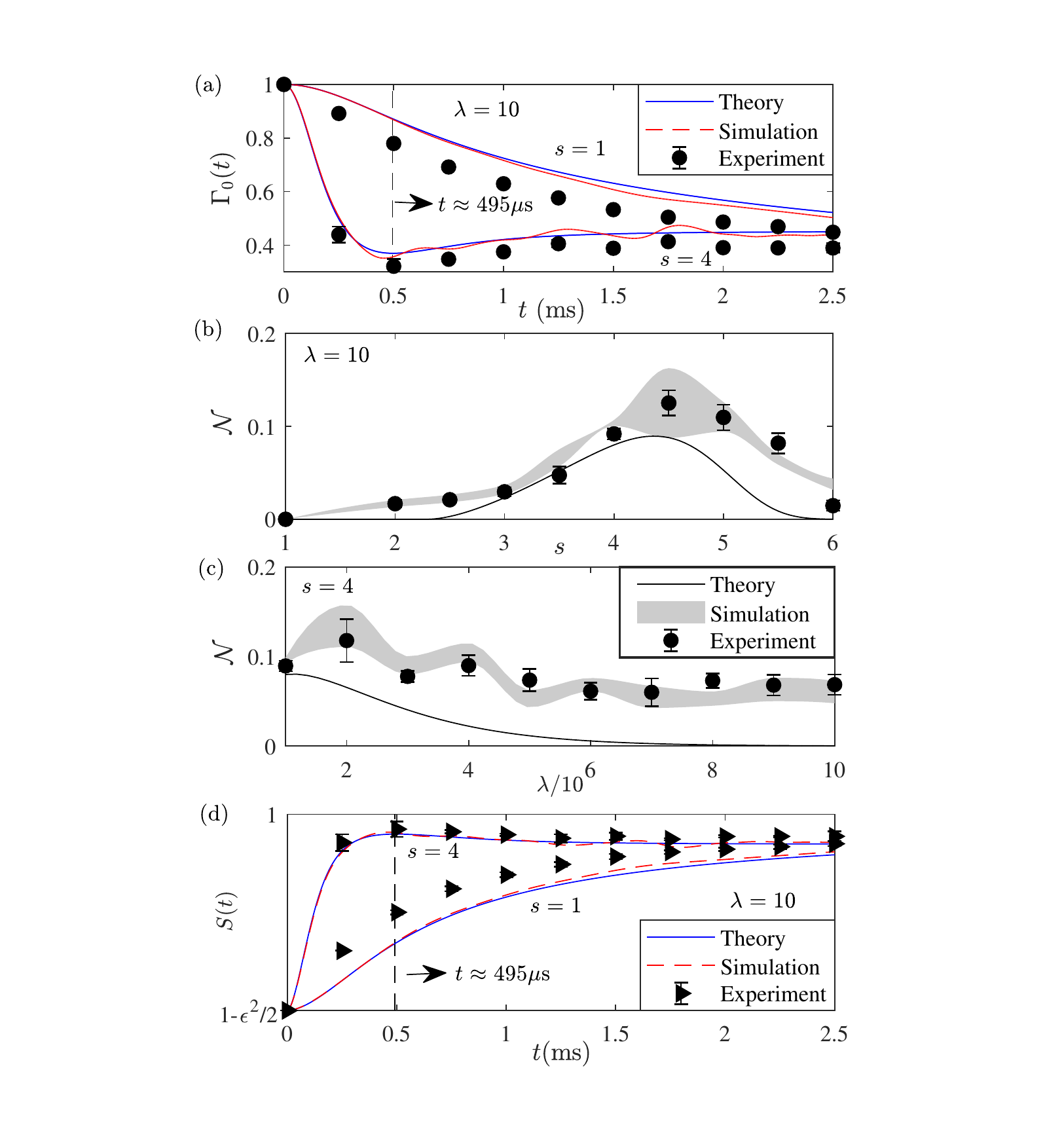}
%	\vspace{-0.3cm}
	\caption{(a) Temporally averaged decoherence functions for BLP Markovian ($s=1, \lambda = 10$) and BLP non-Markovian ($s=4,\lambda = 10$) emulated environments.   Non-Markovianity measure $\mathcal{N}$ versus (b) Ohmicity parameter $s$ for $\lambda = 10$  and (c) coupling constant $\lambda$ for $s = 4$. The shaded regions for simulations and error bars in experiments correspond to standard deviations over 10 distinct bins each of 900 realizations, and they capture the finite-ensemble effects. (d) Variation of von-Neumann entropy of the system with time. }
	\label{NM}
%	\vspace{-0.7cm}
\end{figure} 
We tuned the strength of injected noise $\lambda \in [10,100]$, the base and cut-off frequencies $\omega_b = 2\pi \times 4 $ rad/s and $\omega_c = 2\pi \times 320 $ rad/s, respectively, and the Ohmicity parameter $s\in[1,6]$ so that the signal decays out in 2.5 ms ($\approx 5\ \omega_c^{-1}$).    Figure \ref{NM}(a) contrasts the  temporally averaged signal in the presence of the  BLP Markovian ($s=1,\lambda=10, \mathcal{N} = 0$) environment with the BLP non-Markovian environment  ($s=4,\lambda=10, \mathcal{N}> 0$) with corresponding theory and numerical simulations.  For the non-Markovian case, the onset of information backflow ($\dot{\Gamma}_0(t) > 0$) occurs at about 495 $\mu$s as marked by the dashed line.
In our experiments, since $\Gamma_0(t)$ decays out in $2.5$ ms $\ll T_2$, the intrinsic transverse dephasing has little effect on the engineered dephasing. However, the discrepancy between the simulations and experiments is mainly due to other experimental limitations such as spatial inhomogeneity in RF pulses.
The experimentally obtained non-Markovianity measure $\mathcal{N}$ versus Ohmicity parameter $s$ (Fig. \ref{NM}(b)) and coupling constant $\lambda$ (Fig. \ref{NM}(c)) show an overall agreement with the corresponding theoretical \cite{addis2014comparative,guarnieri2014quantum} and numerical simulations.  
For free evolution (Eq. \ref{chicl}), the non-Markovianity measure is almost zero for low values of the Ohmicity  parameter ($s \leq 2$) as well as for high values ($s>6$), due to minimal overlap of the spectral density $S(\omega)$ with the high-frequency parts of the filter function $\abs{F_0(\omega,t)}^2$.
For $\lambda=10$, the maximum non-Markovianity is obtained for $s=4.5$.  
Unlike the theoretical curves, the finite size of the temporal ensembles ($N=1000$) leads to  additional oscillations in experimental as well as simulated curves (Fig. \ref{NM}(a)), resulting in an overestimation of the non-Markovianity parameter $\cal{N}$ in Fig. \ref{NM}(b) and (c).  To this end, an appropriate smoothening procedure was adopted to minimize the effect of such spurious signal oscillations (see Appendix).
%\begin{figure}
%	\includegraphics[trim = 5cm 12cm 5.9cm 9cm, clip, width=7.5cm ]{Von_neumann_entropy_vs_NM.pdf}
%	\caption{Variation of von Neumann entropy with time. For non-Markovian environment ($s=4$), a decrease in entropy  is observed after \textcolor{red}{$t = t_c= 1/\omega_c \approx 495 \mu$s} in contrast to monotonic increase in case \textcolor{red}{of} Markovian environment $(s = 1)$.   }
%	\label{Ent}
%\end{figure} 
It is also interesting to look at the von-Neumann entropy of the system  $\mathcal{S}(t)$ which provides a thermodynamic perspective on the transition from Markovian to non-Markovian dynamics.  For a single qubit with an initial state $\rho = \mathbbm{1}/2 + \epsilon \sigma_x/2$, 
\begin{equation}
\mathcal{S}(t) = -\tr[\rho(t) \log_2 \rho (t)] \approx 1-\epsilon^2 \Gamma_0^2(t)/2,
\end{equation}
(blue line), along with corresponding simulated (red line) and experimental (symbols) entropies, is shown in Fig.  \ref{NM} (d).  While the monotonic growth of entropy for $s=1$ indicates Markovian behavior, the slight drop of entropy from $t\approx 0.5$ ms for $s=4$ is a signature of non-Markovianity.

\section{DD for non-Markovian environment}

\begin{figure}
	\subfigure{\includegraphics[trim = 2.3cm 10cm 2.3cm 10cm, clip, width=9cm ]{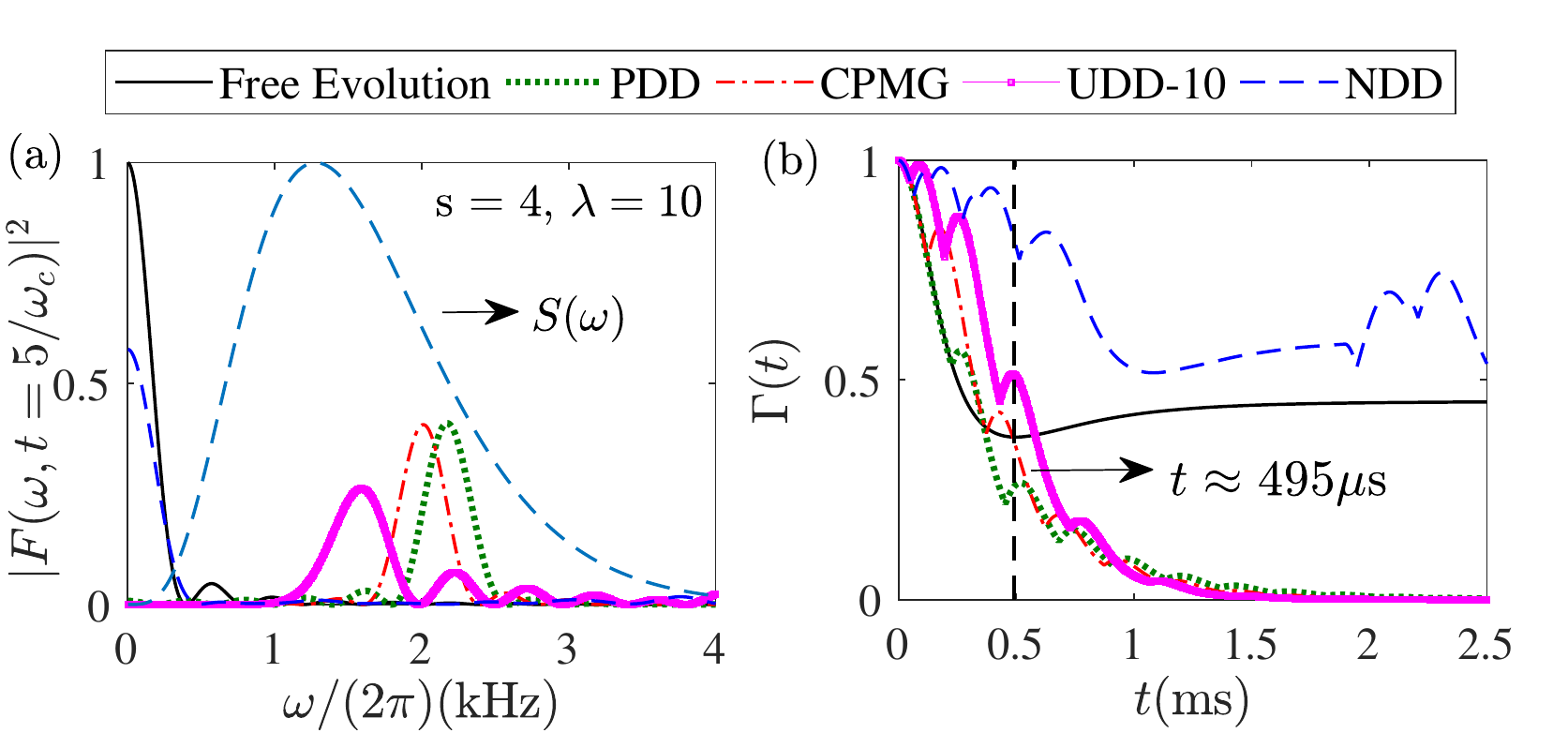}}
	\subfigure{
		\includegraphics[trim = 4.3cm 5.5cm 3.8cm 5.8cm, clip, width=8.5cm ]{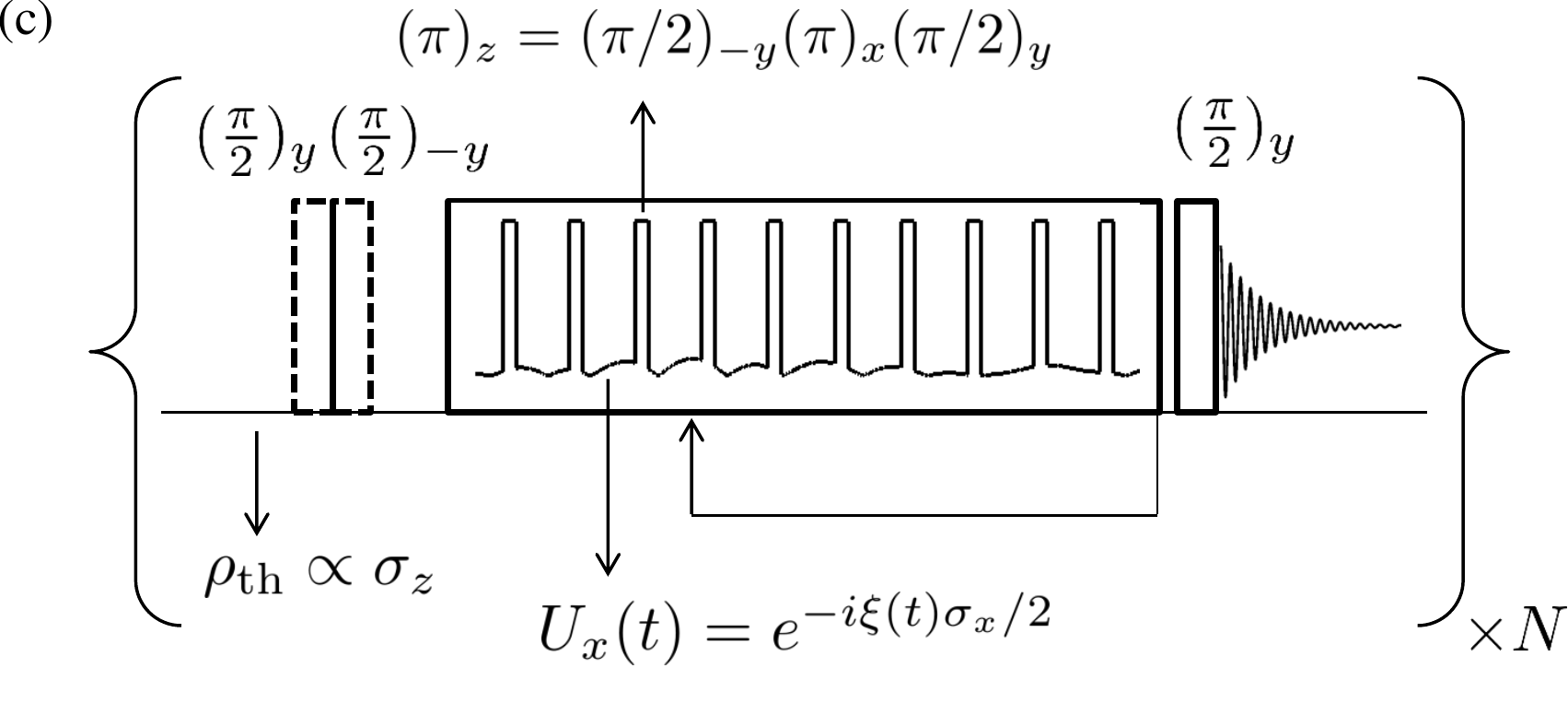}}
	\caption{Noise spectrum (dashed line) for BLP non-Markovian environment corresponding to $s=4,\ \lambda = 10$ and filter-functions $\vert F(\omega,t)\vert^2$ at $t=5\ \omega_c^{-1}$ for various DD sequences as well as for free evolution (a) and corresponding decoherence functions (b).  (c) Circuit used to apply dynamical decoupling along with noise modulation.}
	\label{fig3}
\end{figure}

 Consider a DD-protected qubit undergoing sequential phase-flips ($\pi$ pulses) represented by the rectangular-wave modulation function $f(t) \in \{-1,1\}$. In this case, the effective decoherence function  $\Gamma (t) = e^{-\chi^c (t)}$, where $\chi^c (t)$ has a similar form to that in Eq. \ref{chicl}, except that the filter function is replaced with the Fourier transform,
\begin{eqnarray}
F(\omega,t) =  \int_{0}^{t}
f(t') e^{-i\omega t'} dt'.
\end{eqnarray}
Construction of a DD sequence is based on engineering a filter function $F(\omega,t)$ which minimizes its overlap with a given noise power spectrum $S(\omega)$, and thereby minimizes $\Gamma(t)$.  Therefore, the performance of a DD sequence depends crucially on the timescale associated with the environmental correlation function. Only in the case of $s = 0$ for very large value of $\omega_c$ (white noise), decay of coherence is exponential and DD sequences generally fail. In the presence of BLP Markovian environments ($s\leq2$, $\mathcal{N} = 0$), it has been theoretically shown that the  periodic dynamical decoupling (PDD) sequence \cite{viola1999dynamical} is most efficient  \cite{addis2015dynamical} when delay between inversion pulses $\Delta t$ is smaller than $\omega_c^{-1}$ as expected. However, the PDD sequence becomes inefficient as non-Markovian effects become relevant ($s>2$, $\mathcal{N} > 0$) even when $\Delta t< \omega_c^{-1}$. This can be easily understood in terms of filter function formalism as shown in Fig \ref{fig3}(a) for $\Delta t \approx 0.5\ \omega_c^{-1}$ in the presence of the BLP non-Markovian environment corresponding to the Ohmicity parameter $s=4$ and the coupling constant $\lambda=10$. Along with PDD, we also plot the filter function for CPMG \cite{carr1954effects,meiboom1958modified} and  Uhrig dynamical decoupling (UDD) \cite{uhrig2007keeping} sequences. Due to substantial overlap of the noise spectrum with the filter function, these sequences under perform for $s>2$. 

If we pack more $\pi$ pulses in a given total duration, CPMG and PDD filter function peaks will move to higher frequencies, thus reducing the overlap with the noise spectrum. However for a fair  comparison, we keep the same number of $\pi$ pulses over a fixed total duration for all DD sequences. Each $\pi$ pulse, instead of being an instantaneous spin flip,  has in practice a finite duration, associated pulse errors, and requires certain energy output from the duty-cycle limited hardware \cite{souza2012robust,souza2011robust,ahmed2013robustness}. For a given duration, the total number of $\pi$ pulses therefore constitutes the resource required for DD.

 We synthesize the \textit{non-Markovian DD-sequence} (NDD) that maximizes the coherence protection parameter $\mathcal{P} = \int_{0}^{t} dt'\Gamma (t')/t $ \cite{addis2015dynamical} for a known non-Markovian environment by numerically optimizing the time-instants of $\pi$ pulses keeping the total number of pulses ($n$) constant. Specifically, we use a genetic algorithm for optimization in this $n$-dimensional space where genes are delay between the $\pi$ pulses. We provide CPMG as an initial guess to the algorithm and optimization was constrained by the minimum delay between DD pulses, i.e., the																																																																																																																																																																																																																																																																																																																																															width of the DD pulse itself, which was $50\mu$s. The algorithm took up to 500 generations to reach an optimal solution. The circuit used to apply DD pulses along with noisy modulation is shown in Fig
 \ref{fig3} (c).  We interleave $\pi_z$ pulses each of duration $50\mu$s with the noise modulation profile for each realization and the noisy modulation is only applied during free-precession time to maintain high  fidelity of $\pi_z$ pulses.
  
  \begin{figure}
  	\subfigure{\includegraphics[trim = 2.2cm 8cm 2.1cm 7.8cm, clip, width=8.7cm ]{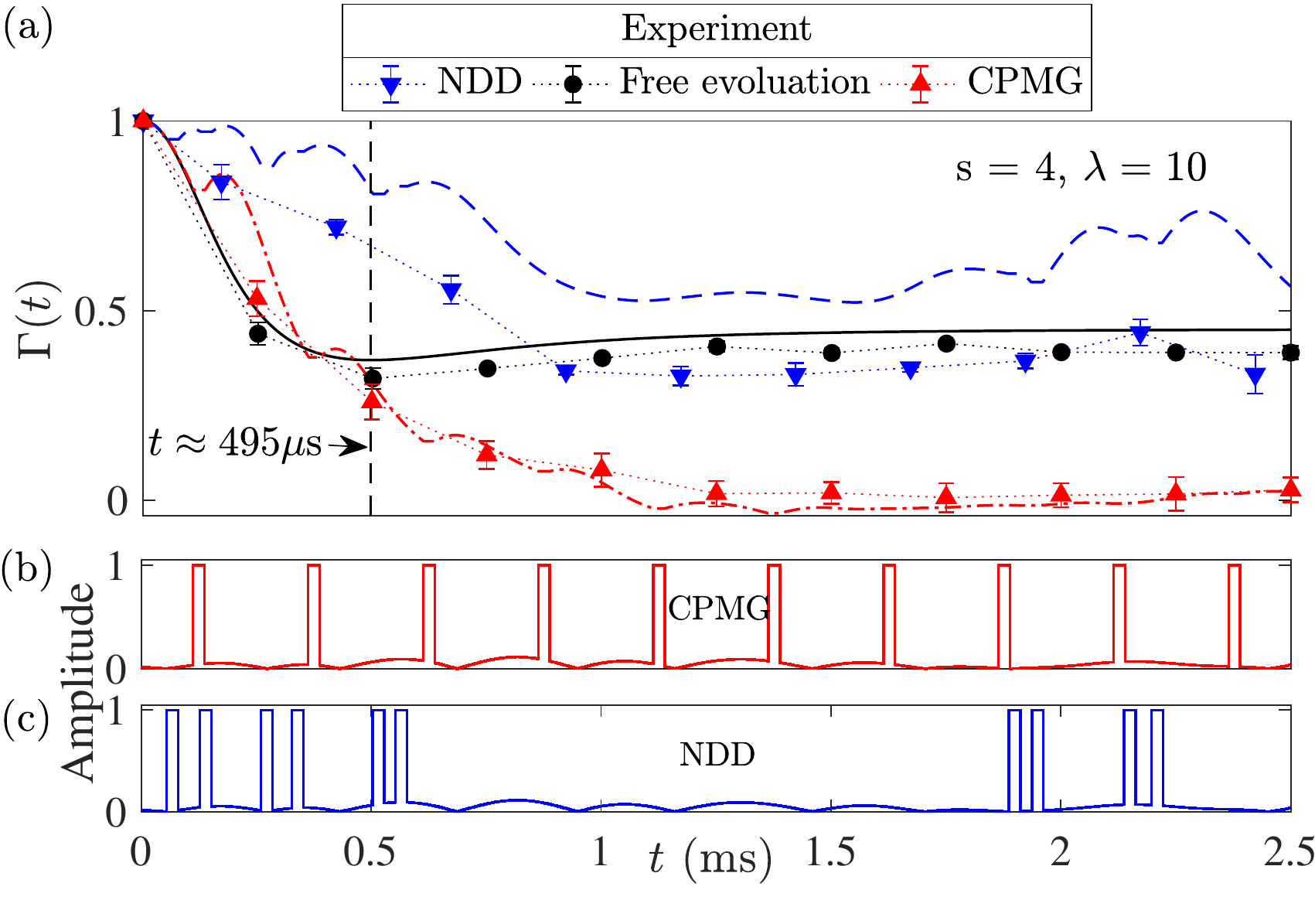}}
  	
  		\subfigure{\includegraphics[trim = 1.8cm 8cm 1.2cm 7.6cm, clip, width=8.7cm ]{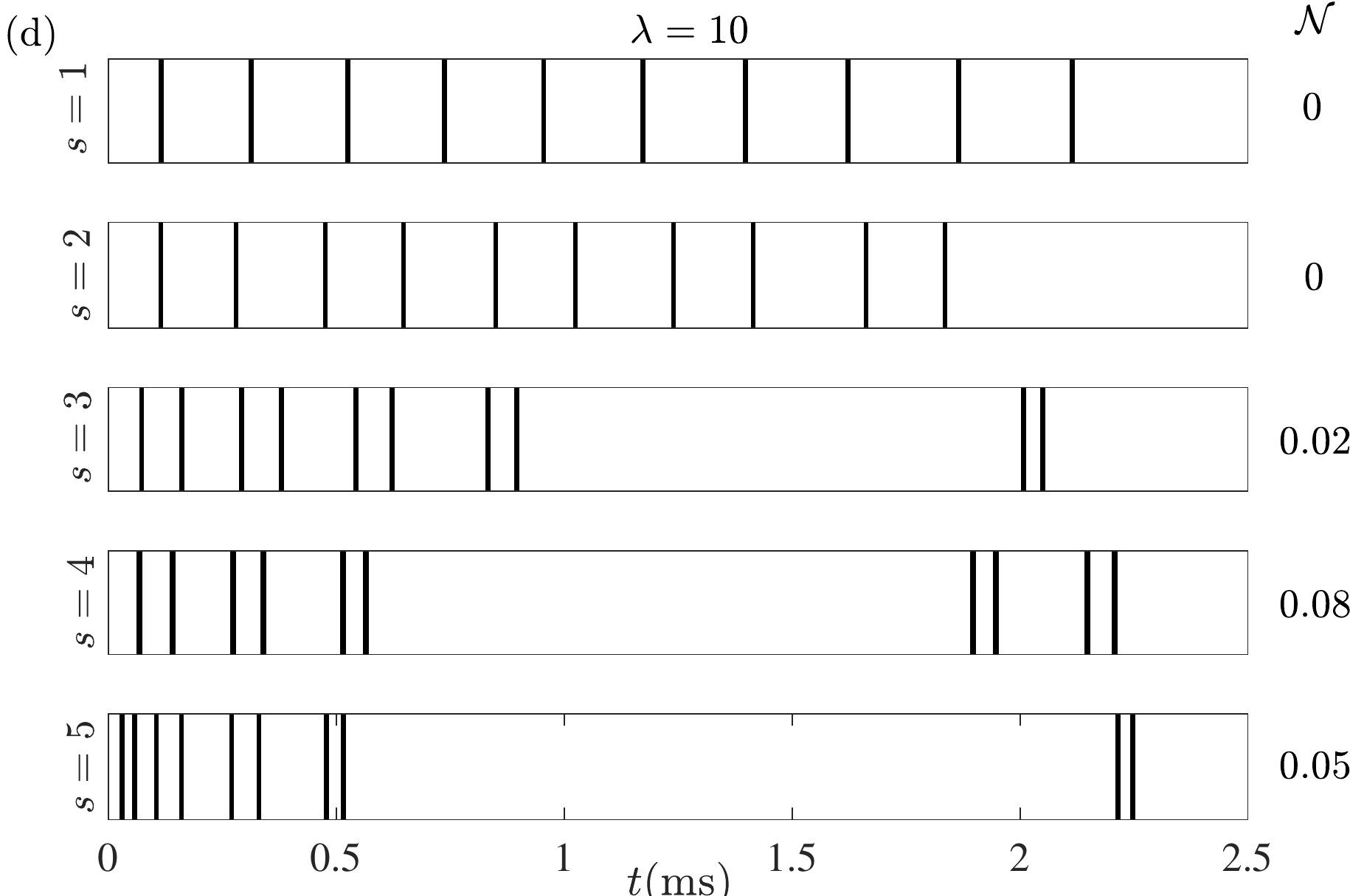}}
  	\caption{
  		(a) Theoretical  and experimental (symbols) decoherence functions with free evolution (smooth), CPMG (dash-dotted) (b), or NDD (dashed) (c) sequences. In (b) and (c) one particular noise realization is interleaved with $\pi$ pulses. (d) Optimal DD sequences for various Ohmicity parameter $s\in\{1,5\}$. Corresponding values of non-Markovianity measure are also indicated. }
  	
  	\label{DDcomp}
  \end{figure} 
 Filter functions for the NDD sequence designed for the BLP non-Markovian environment ($\mathcal{N} > 0$) corresponding to Ohmicity $s = 4$ with noise strength $\lambda = 10$ is shown in Fig. \ref{fig3} (a) along with other sequences including free evolution.  Note that the NDD filter function has the minimal overlap with $S(\omega)$ indicating a better coherence protection as evident from the corresponding decoherence functions plotted in Fig. \ref{fig3} (b).  The inefficiency of other sequences can be attributed to the localization of the noise strength at the intermediate frequencies. Figure \ref{DDcomp} (a) compares the theoretical and experimental (symbols) performances of the CPMG sequence (Fig. \ref{DDcomp} (b)) with NDD (Fig. \ref{DDcomp} (c)) and free evolution for the BLP non-Markovian environment corresponding to  $s = 4$ and $\lambda = 10$. Each of the 1000 realizations for emulated environmental $x$ modulations (see Fig. \ref{circuit}) was interleaved with a total of ten composite $\pi_z$ pulses of width $50~\upmu$s . Experimentally, the fidelity of $\pi_z$ pulse was $\approx 98\%$ which indicates towards $15\%$ radio-frequency inhomogeneity. It is interesting to note that for durations less than 0.5 ms, CPMG shows faint improvement over free-evolution.  However, once the information back-flow sets in, CPMG not only fails to protect the coherence, but also has a detrimental impact on it.  In contrast, the NDD sequence should have a much better coherence protection as indicated by the theoretical decoherence function.  Experimentally, there is a significant protection for up to 1 ms, and then the performance drops below free-evolution presumably due to finite pulse-widths, calibration errors, and other pulse-imperfections.  Generation of numerous NDD sequences starting from random guesses and for various values of Ohmicity parameter $s\in \{1,5\}$, revealed a general pattern involving bunching of $\pi$ pulses at the beginning and at the end of  sequences for BLP non-Markovian environments $(s>2)$.  This feature explains the exclusion of the filter function at the intermediate frequencies as observed in Fig. \ref{DDcomp}(a) and consequently the minimization of overlap with the spectral density.  This pattern may help the NDD sequence to take advantage of the inherent information backflow by avoiding $\pi$ pulses in the intermediate time durations.

 \begin{figure}
	\includegraphics[trim = 1.6cm 7.7cm 2.2cm 7.8cm, clip, width=8.65cm ]{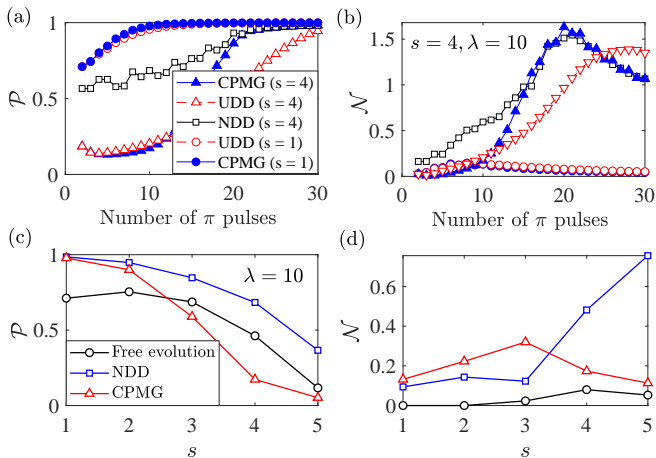}
	\caption{Numerically evaluated coherence protection ${\cal P}$ and non-Markovianity parameter ${\cal N}$ is plotted for various DD sequences versus  number of $\pi$ pulses in (a) and (b). Same parameter plotted for various values of $s$ under the application of CPMG, NDD, as well as free evolution. NDD sequences shown in Fig. \ref{DDcomp} (d) were used in this case. }
	\label{fig5}
\end{figure}
In  Fig. \ref{fig5}(a) and Fig. \ref{fig5}(b), the protection parameter $\mathcal{P}$ and non-Markovianity measure $\mathcal{N}$, respectively, are plotted versus the number of $\pi_z$ pulses (various $\Delta t$ regimes) for CPMG, UDD, and optimal NDD sequences. As expected, in the absence of information backflow ($s=1, \lambda = 10$), the protection parameter increases linearly before saturation, in contrast to a nonlinear behavior (before saturation) in the presence of information backflow ($s=4, \lambda = 10$). For high value of the number of pulses (for shorter $\Delta t$) the optimal solution reaches CPMG irrespective of noise spectrum (Markovian or non-Markovian).   
Interestingly, independently of noise spectra ($s = 1$ or $s = 4$) and the DD sequence used, the non-Markovianity measure increases with number of $\pi$ pulses and starts decreasing after the protection parameter achieves the maximum value. This behavior indicates that the number of spin flips can be regarded as a tuning knob to engineer the desired non-Markovian environments.

Since the concept of DD is also associated with information backflow, it is natural to ask the question of 	whether there is any correlation between protection provided by optimal DD sequences and non-Markovianity introduced by them in terms of BLP measure. In this regard, we plot parameter $\mathcal{P}$  as well as $\mathcal{N}$ for CPMG and  NDD sequences as a function of Ohmicity parameter $s$ for $\Delta t = 0.5\ \omega_c^{-1}$ and $\lambda = 10$ in Fig. \ref{fig5} (c) and (d). Non-Markovianity is higher compared to free evolution for both CPMG and NDD sequences. However, we do not observe any clear evidence of correlation between protection and non-Markovianity.       

\section{Conclusion}
 We described  experimentally emulating the non-Markovian dynamics of a pure dephasing spin-boson model at zero temperature by engineering a noise power spectrum with the help of a temporally averaged set of randomized external fields. We characterized the emulated non-Markovianity using the BLP measure \cite{breuer2009measure} and von Neumann entropy of the system. Emulating quantum non-Markovian dynamics is important not only from the fundamental point of view  to understand dynamics of information backflow \cite{breuer2009measure,breuer2016colloquium,de2017dynamics,guarnieri2014quantum,liu2011experimental,rivas2010entanglement,rivas2014quantum} and thermodynamic properties such as flow of heat and  entropy production \cite{vinjanampathy2016quantum,jarzynski2017stochastic,strasberg2016nonequilibrium,katz2016quantum,friedman2018quantum},  but also from the practical perspective of developing coherence protection protocols in the presence of environmental memory effects \cite{addis2015dynamical,mukherjee2015efficiency,reich2015exploiting,tai2014optimal,addis2016problem}. With excellent control over quantum dynamics, we believe that  NMR systems are excellent experimental test beds to study more complex and not exactly solvable non-Markovian dynamics. 
 
 We experimentally investigated the efficiency of the CPMG DD sequence in the presence of non-Markovian environments with nonzero BLP measure. Moreover, using the filter function formalism \cite{uhrig2008exact,biercuk2009optimized,cywinski2008enhance,gordon2008optimal,uys2009optimized,clausen2010bath} we designed  DD sequences that optimize the position of $\pi$ pulses (phase flips) to maximize coherence protection for a specific non-Markovian environment. We observed a bunching of $\pi$ pulses at the beginning and end of the sequence which hints towards exploiting information backflow associated with non-Markovianity of dynamics.   This pattern might be insightful to incorporate non-Markovianity into the optimization routines. As the number of $\pi$ pulses is increased keeping the total duration fixed, the optimal sequences approach toward the CPMG sequence as expected. The BLP measure increases with number of $\pi$ pulses till maximum protection is achieved. This aspect can be used as a tuning knob to engineer non-Markovianity in a systematic fashion.  We believe that our investigations constitute an important step  to study the impact of memory effects of environment on more involved quantum control protocols and contribute towards understanding non-Markovianity as a resource for quantum technologies.      

\section*{Acknowledgments}
 This work was supported by DST/SJF/PSA-03/2012-13 and CSIR 03(1345)/16/EMR-II.
 \vspace{0.4cm}
\section*{APPENDIX}
  \label{app1}
\subsection*{Estimation of non-Markovianity measure $\mathcal{N}$}

   \begin{figure}
  	\includegraphics[trim = 1.3cm 6.9cm 1.2cm 7.2cm, clip, width=8.6cm ]{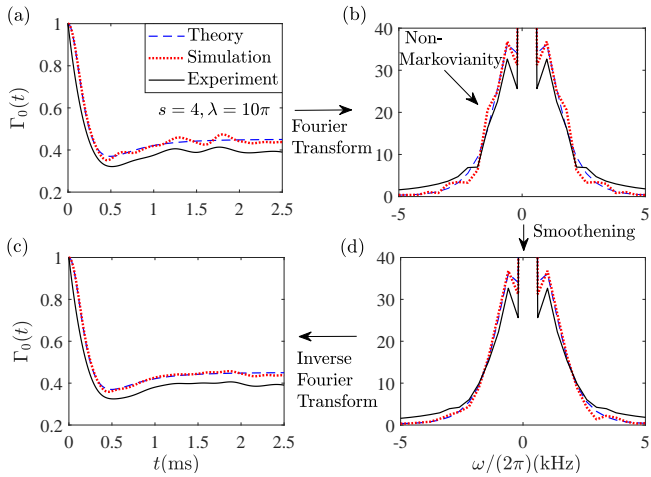}
  	\caption{Smoothening process to estimate non-Markovianity measure $\mathcal{N}$}
  	\label{smoothning_process}
  \end{figure}  
  For emulating non-Markovian dynamics by noise power spectrum engineering using randomized control fields, a finite number of realizations are possible due to experimental constraints of time required (five times longitudinal relaxation time $5\times T_1 = 2$s) for reinitialization after every realization of $\xi(t)$. This produces oscillation artifact on top of the characteristic nonmonotonicity of the decoherence function due to non-Markovianity [Fig \ref{smoothning_process}(a)]. Smoothening can not be used on time domain data directly because it can not differentiate between spurious oscillations and the concerned non monotonicity. However, in the Fourier domain these two can be separated since the artifact appears as noise on top of the peak due to non-Markovianity in the Fourier transform of $\Gamma_0(t)$ [Fig. \ref{smoothning_process}(b)]. We smoothen out these oscillation using standard data processing techniques [Fig \ref{smoothning_process}(c)] keeping the maximum of the peak intact and then we inverse Fourier transform  to get the smoothened decoherence function [Fig \ref{smoothning_process}(d)].   
 
 \bibliography{NM}{}
%    \vspace{-0.4cm}
\bibliographystyle{apsrev4-1}

\end{document}